\documentclass[12pt,a4paper]{article}
\usepackage{amsmath,amscd}
\usepackage{graphicx}
\usepackage[usenames,dvipsnames]{xcolor}
\usepackage{mathpazo} % typeface
\usepackage{hyperref}
\usepackage{multimedia}
\usepackage{comment}
\usepackage[small,nohug,heads=vee]{diagrams}
\diagramstyle[labelstyle=\scriptstyle]
\definecolor{light-gray}{gray}{0.80}

\usepackage{fancyhdr}
\renewcommand{\maketitle}{
    \begin{center}
      \Large
        {\bf The Lorentz Force and Energy-Momentum for \\ Off-Shell Electromagnetism}
        \vskip .3 true cm
      \small
	M.C. Land$^{1}$ and L.P. Horwitz$^{2}$ \\
	School of Physics and Astronomy \\
	Raymond and Beverly Sackler Faculty of Exact Sciences \\
	Tel Aviv University, Ramat Aviv, Israel  \\
email: $^{1}$martin@hadassah.ac.il, $^{2}$larry@post.tau.ac.il
      \end{center}
      \vskip .5 true cm
}
\begin{document}
\title{}
\author{}
\begin{flushright}
TAUP 1824-90
\end{flushright}
\vskip 3 true cm
\maketitle
\begin{abstract}
The kinematics of pre-Maxwell electrodynamics is
examined and interpretations of these fields is found through an examination
of the associated Lorentz force and the structure of the energy-momentum
tensor.  
\vskip 0.5 true cm

{\noindent}Key words: relativistic quantum theory, off-shell gauge fields,
pre-Maxwell fields, higher dimensional gauge fields.
\end{abstract}

\baselineskip7mm \parindent=0cm \parskip=10pt

%\vskip 0.5 true cm
\par In the framework of a covariant relativistic quantum theory [1][2], in
which the evolution of the system is parametrized by a universal O(3,1)
invariant world (or historical) time $\tau$, Saad, Horwitz and Arshansky [3]
have shown that the requirement of local gauge invariance leads to five
compensation fields.  These fields, which have been called pre-Maxwell fields,
are defined on a five dimensional manifold with coordinates $(x^\mu, \tau)$,
and the gauge symmetry of the equations of motion is associated with a five
component conserved current consisting of an O(3,1) four-vector current, $
j^\mu$, and a scalar density (in $R^4$), $\rho$. The field equations, obtained
from a Lagrangian constructed from gauge invariant field strengths, have a
five dimensional symmetry which could be O(3,2) or O(4,1).  The propagators
for these field equations, classified according to their spacetime asymptotic
properties, have been worked out in ref. [4].
  \par In this letter we shall examine the kinematics of the pre-Maxwell
electrodynamics and find interpretations of the new fields through an
examination of the Lorentz force and the structure of the energy-momentum
tensor of the field.
  \par The evolution equation [1] (we use the metric $-+++$ for the O(3,1)
indices)
$$i {\partial\over\partial\tau} \psi_\tau = {p^\mu p_\mu \over 2M} \psi_\tau
\eqno(1) $$
may be made locally gauge invariant under the transformations (we shall use $x
\equiv x^\mu$)
$$ \psi \rightarrow e^{ie_0 \Lambda(x,\tau)} \psi \eqno(2) $$
through the introduction of compensation fields $a_\alpha= (a_\mu,a_5)$ which
transform as $(\partial_\alpha=\partial/\partial{x^\alpha},dx^5=d\tau)$
$$ a_\alpha (x,\tau) \rightarrow a_\alpha (x,\tau)+\partial_\alpha
\Lambda(x,\tau) \eqno(3) $$
   \par The gauge invariant evolution equation then becomes
$$ i{\partial\over \partial\tau} \psi_\tau (x,\tau) = [{1\over2M}(p_\mu -e_0
a_\mu)(p^\mu -e_0 a^\mu) -e_0 a_5] \psi_\tau (x). \eqno(4) $$
It follows from this equation (as for the case of $\tau$-independent Maxwell
fields treated by Stueckelberg [1]) that
$$ \partial_\mu j^\mu +\partial_\tau \rho =0 \eqno(5)$$
i.e.,
$$ \partial_\alpha j^\alpha =0 \eqno(6) $$
where
$$ j_\tau^5 = \rho = |\psi_\tau (x)|^2 $$
and
$$ j_\tau^\mu = -{i\over2M}[\psi_\tau^* (\partial_\mu -ie_0 a^\mu )\psi_\tau
-\psi_\tau (\partial_\mu +ie_0 a^\mu )\psi_\tau^* ] \eqno(7) $$
  \par We shall, in the following, consider the right hand side of (4) as
providing the form of the classical relativistic generator of
$\tau$-evolution, i.e., the Hamiltonian K for the classical mechanics of
events in spacetime.  The canonical Hamilton equations
$$ {dx^\mu \over d\tau} = {\partial K\over \partial p_\mu } = {p^\mu -e_0
a^\mu\over M} $$
enable us to write the Lagrangian (we shall write four vector products as
e.g., $p^2 \equiv p^\mu p_\mu , x\cdot p \equiv x^\mu p_\mu $)
$$ L= {dx \over d\tau} \cdot p -K $$
as
$$ L= {M \over 2}({dx\over d\tau })^2 +e_0 (a_\alpha {dx^\alpha\over d\tau }),
\eqno(8) $$
where we have used the fact that $dx^4 /d\tau = 1$.  To provide the pre-
Maxwell fields with dynamical structure, we introduce a kinetic term for the
field to the Lagrangian. The total action is then
$$ S=\int d\tau {M\over 2}({dx\over d\tau})^2 +e_0 a_\alpha {dx^\alpha \over
d\tau} -{\lambda \over 4}\int d\tau d^4 x' f^{\alpha\beta}
(x',\tau)f_{\alpha\beta} (x',\tau), \eqno(9) $$
where
$$ f_{\alpha\beta} =\partial_\alpha a_\beta - \partial_\beta a_\alpha
\eqno(10) $$
To raise and lower the fifth index $\alpha =5$, we use a formal metric
signature that we shall call $\sigma =\pm $ (to be specified later), i.e., our
five dimensional signature is of the form ($-,+,+,+,\sigma$). The first,
O(3,1) invariant, term of the Lagrangian, associated with the matter
equations, breaks the O(3,2) or O(4,1) symmetry corresponding to this metric.
  \par The variation of this action with respect to the path $x(\tau)$ leads
to the equation of motion
$$ M {d^2 x^\mu \over d\tau^2} = e_0 f^{\mu\alpha}{dx_\alpha \over d\tau} =
e_0 (f^{\mu\nu} {dx_\nu \over d\tau} -\sigma f^{5\mu} ), \eqno(11) $$
the Lorentz force associated with the pre-Maxwell fields.
  \par As for the usual Maxwell analog, we calculate the derivative of the
kinetic term for the event motion:
$$ {d\over d\tau} [{1\over 2}M({dx\over d\tau })^2]= M({dx\over
d\tau})\cdot({d^2 x\over d\tau ^2}) = -\sigma e_0 f^{5\mu} {dx_\mu \over
d\tau} , \eqno(12) $$
where we have used (11) and the antisymmetry of $f^{\mu\nu}$. Hence we see
with the help of (12) that one obtains the concise form
$$ M {d \over d\tau} (p^\alpha -e_0 a^\alpha ) = e_0 f^{\alpha\beta} {dx_\beta
\over d\tau} \eqno(13) $$
for a formal ``five vector'' $p_\alpha = (p_\mu, K_{kin})$, where $K_{kin}$ is
the kinetic part of K.  This result is in the form of a generalized Lorentz
force  which is formally covariant over a larger group with the signature
($-,+,+,+,\sigma$), i.e., O(3,2) or O(4,1).  We note that the Lorentz force is
proportional to $e_0$, not the dimensionless Maxwell charge e (the field
strengths $f^{\alpha\beta}$ have dimension ${1\over L^4 }$).
\par   We remark that the integral of (12) over $\tau$ results in
$$ {1\over 2M}(m^2 (\infty) -m^2 (-\infty) ) =\sigma e_0
\int_{-\infty}^{\infty} d\tau f^{5\mu} ({dx_\mu \over d\tau}). \eqno(14) $$
where $m^2 (x)=-(p-e_0 a)^2$.  The four-vector $f^{5\mu}$ is seen here to
generate changes in the Lorentz invariant mass squared.  Asymptotic
conservation of mass would imply a systematic cancellation on the right hand
side side, for example, through oscillations in $dx^\mu /d\tau$ or for
$f^{5\mu}$ spacelike at $x^\mu (\tau)$ for each $\tau$.   The mass squared
may, however, undergo local dynamically induced variations even if it is
asymptotically conserved.
   \par The phenomenon of (classical) pair annihilation or production [1]
requires such a mass change; since $K=-(M/2)(ds^2/d\tau^2) -e_0 a_5 $, a
change in sign of $ds^2 = dt^2 - d{\bf x}^2$ may occur when compensated by
$a_5$.  In the Maxwell analog of the extension of (11) to (13), the derivative
of the usual kinetic energy is proportional to $F^{0j} v_j (= {\bf E} \cdot
{\bf v})$, where $v_j = dx_j /dt $, and $F^{\mu\nu}$ is the Maxwell field, and
vanishes when ${\bf v}$ does.  Both for relativistic and nonrelativistic
kinematics, vanishing of the kinetic energy implies that ${\bf v} = 0$, and
hence no further decrease (to negative values, corresponding to a turning back
of the trajectory) is possible.  In Eq. (12), it is clear that vanishing of
the {\it mass} (squared), i.e., $ ({dx\over d\tau })^2$, does not imply that
$dx^\mu /d\tau$ vanishes.  Hence, it is possible for the motion to pass
through the light cone, resulting in possible pair annihilation.  Stueckelberg
[5] found that although classical pair creation and annihilation can, in
principle, occur in the framework of the manifestly covariant  theory, such
processes are not induced by standard Maxwell fields (the sign of the
invariant interval is conserved). We see, in the analysis of (11) and (12),
that $\tau$-dependence of the fields is essential (so that $f^{5\mu}$ is not
zero). We remark that, in the      pre-Maxwell theory, moreover, if ${dx^\mu /
d\tau} = 0$, the $\alpha=5$ term contributes to $d^2 x^\mu /d\tau^2$ the
quantity $e_0 f^{\mu 5}$; hence the sign of $dx^0 / d\tau$  can change under
the influence of the $f^{\mu 5}$ field. An example of this phenomenon is given
in the next section.
  \par The variation of the action with respect to the fields $a_\alpha
(x,\tau)$ leads to the field equations
$$ \partial_\beta f^{\alpha\beta} (x,\tau) =(e_0 /\lambda)({d\over d\tau}
x^\alpha (\tau) ) \delta^4 (x-x(\tau)) \equiv ej^\alpha (x,\tau) \eqno(15) $$
where $e \equiv e_0 /\lambda$.  Integrating the $\alpha = \mu$ components of
this equation on $(-\infty,\infty)$ over $\tau$, with the condition that
$f^{\mu 5}$ vanishes pointwise at $\tau \rightarrow \pm \infty$, we recover
the Maxwell equation [3][4]
$$ \partial_\nu F^{\mu\nu} (x) = eJ^\mu (x) \eqno(16) $$
It follows, as for the quantum case (5), from (this condition is necessary for
the consistency of (15) and the gauge invariance of the action) [1][2]
$$ \partial_\mu j^\mu (x,\tau) + {\partial\rho\over \partial\tau} = \partial_
\alpha j^\alpha (x,\tau) =0 \eqno(17) $$
where $\rho = \delta^4 (x-x(\tau)) $, that
$$ J^\mu (x) = \int d\tau j^\mu (x,\tau) \eqno(18) $$
is conserved if the density $\rho$ vanishes pointwise at $\tau \rightarrow \pm
\infty$.  This integration for $J^\mu$ has been called ``concatenation'' [6],
and provides the link between the notion of an {\it event} along a world line
and the notion of a {\it particle}, whose support in spacetime is the whole
world line.  We see from (16) that the Maxwell potential is given by the
concatenation of the pre-Maxwell field
$$ A^\mu (x) = \int d\tau a^\mu (x,\tau) \eqno(19) $$
and that the constants $e_0$ and $\lambda$ have the dimension of length. Their
ratio is identified by (16) as the dimensionless electric charge.  The
representation of $f_{\alpha\beta}$ as the antisymmetric derivative of a five-
field is equivalent to the homogeneous equation
$$ \partial^\alpha \epsilon_{\alpha\beta\gamma\delta\sigma} f^{\delta\sigma}
=0, \eqno(20) $$
analogous to the homogeneous Maxwell equations.
 \par  We now examine the noncovariant form of the field equations [3].  Let
us define the three-vectors
$$ e_i =f^{0i} \qquad h_i = {1\over 2} \epsilon_{ijk} f^{jk} , \eqno(21) $$
and the four-vector $\epsilon^\mu$ defined by
$$ \epsilon^i = f^{5i} \qquad \epsilon^0 =f^{50} . \eqno(22) $$
The pre-Maxwell equations then become
$$ \nabla \cdot {\mathbf e} = ej^0 + \partial_\tau \epsilon^0 \qquad   \nabla
\times {\mathbf e} + \partial_0 {\mathbf h} =0 $$
$$ \nabla \times {\mathbf h} - \partial_0 {\mathbf e} -\partial_\tau
\boldsymbol{\epsilon}
=e{\mathbf j} \qquad   \nabla \cdot {\mathbf h} =0 $$
$$ \nabla \cdot \boldsymbol{\epsilon} = ej^5 - \partial_0 \epsilon^0 \qquad \nabla
\times \boldsymbol{\epsilon} - \sigma \partial_\tau {\mathbf h} =0 $$
$$ \nabla \epsilon^0 = -\sigma \partial_\tau {\mathbf e} -\partial_0 \boldsymbol{\epsilon}  \eqno(23) $$
  In the static limit, for which there is no $\tau$-dependence in the fields,
this system of equations reduces to Maxwell's form for the fields ${\mathbf e}$
and ${\mathbf h}$; the conditions on $\boldsymbol{\epsilon}$ become
$$ \nabla \times \boldsymbol{\epsilon} =0 $$
$$ \nabla \epsilon^0 + \partial_0 \boldsymbol{\epsilon}  =0 $$
$$ \partial_\mu \epsilon^\mu = ej^5 ,\eqno(24) $$
where the last of (24) is generally valid.  From the first of (24), it follows
that there exists a function $\phi (x)$ such that
$$ \boldsymbol{\epsilon} = -\nabla \phi,  \eqno(25) $$
and from the second that $\nabla (\epsilon^0 + \partial^0 \phi )=0$.  We
therefore  may take (up to a constant in $\epsilon^0$)
$$ \epsilon^\mu = -\partial^\mu \phi .\eqno(26) $$
The third of (24) then implies that
$$ -\partial^\mu \partial_\mu \phi = ej^5 = e\rho . \eqno(27) $$
The field $\epsilon^\mu$ is therefore defined by a scalar potential in the
static case; the potential is a Klein-Gordon type massless field whose source
is the (static; e.g., $\delta^4 (x)$) event density.  It is decoupled from the
Maxwell type fields.  The field equations, as for the Maxwell theory, imply
the existence of the vector potential $a^\mu$, for which (10) is valid [3]. 
Since
$$ -\partial^\mu \phi = \epsilon^\mu = f^{5\mu} = -\partial^\mu a^5 ,\eqno(28)
$$
in the static case, we see that (26) is equivalent to (28) when $\phi$ is
identified with $a^5$, the additive potential term in the generator of
evolution K.  As in the nonrelativistic case, we find that the scalar (in this
case Lorentz scalar) component of the gauge field plays the role of a
potential in the static limit. Note that $a_0$ does not change sign under
charge conjugation, but the $a_5$ field does.
\par  We now study the noncovariant form of the Lorentz force.  The space
components of (11) have the form
$$ M \ddot {\bf x} = e_0 [{\bf e} \dot x^0 + \dot {\bf x} \times {\bf h}
-\sigma \boldsymbol{\epsilon} ] \eqno(29) $$
where $\dot x^\mu = dx^\mu /d\tau$.
Since $\dot x^0 =(m/M){1\over \sqrt{1-v^2} } = (m/M)\gamma$ (for $|{\bf v}| =
|d{\bf x} / dt |< 1)$, we may write this equation in the form
$$ {d\over dt} m\gamma {\bf v} =e_0 ({\bf e} + {\bf v} \times {\bf h} -\sigma
\boldsymbol{\epsilon} (M/m\gamma) ) \eqno(30) $$
This equation has the form of the usual Lorentz force, with the additional
contribution of the $\boldsymbol{\epsilon}$ field, which becomes less significant as
$\gamma$ becomes large. Since the left hand side of (30) has a classical
interpretation in terms of motion along the world line, we see that $e_0 ({\bf
e} - \sigma M \boldsymbol{\epsilon} /m\gamma )$ is an effective ``electric'' force. 
The 0th component of (11),
$$ M \ddot x^0 =e_0 ({\bf e} \cdot \dot {\bf x} -\sigma \epsilon^0 ),
\eqno(31)$$
can be written as
$$ {d\over dt} (\gamma m) = e_0 ({\bf e} \cdot {\bf v} -\sigma M \epsilon^0
/m\gamma ); \eqno(32) $$
it represents the work performed on the charge by the field, and is associated
with a change in the ``energy''.  The noncovariant form for the 5 component of
(12) is
$$ {d\over dt} {m^2 \over 2M} = \sigma e_0 (\boldsymbol{\epsilon} \cdot {\bf v}
-\epsilon^0 ); \eqno(33) $$
where we have replaced $(dx/d\tau)^2$ by $-(m^2 /M^2 )$.  We remark that the
magnetic field ${\bf h}$ does not contribute to the change in $\gamma m$ or
the kinetic term $(m^2 /2M)$.  Combining (32) and (33) to eliminate the
$\epsilon^0$ term, one obtains
$$ {d\over dt} (\gamma m) - {1\over \gamma} {dm\over dt} =e_0[{\bf e} -
{\sigma M\over \gamma m} \boldsymbol{\epsilon} ] \cdot {\bf v} , \eqno(34) $$
the linear combination of ${\bf e}$ and $\boldsymbol{\epsilon}$ which, as pointed out
above, corresponds to an effective ``electric'' force.  Taking the scalar
product of  (30) with ${\bf v}$, the term ${\bf v} \times {\bf h}$ drops out,
and we find that $ d/dt (\gamma m) - {1\over \gamma} dm/dt ={\bf v} \cdot d/dt
(m\gamma {\bf v} )$; the term proportional to $dm/dt$ cancels in this
expression, which then reduces to an identity.  This relation reflects the
fact that eq. (33) is not independent of (30) and (32).  We also note that in
the case $dm/dt =0$, it follows from (33) that $\epsilon^0 = \boldsymbol{\epsilon}
\cdot {\bf v}$; (32) then coincides with (34).
\par Let us consider an example in which the world line changes the direction
of its evolution in $t$ due to the presence of the force $-\sigma e_0
\epsilon^0$.  The event enters a region at $\tau$=0 in which $f^{\mu\nu} =
f^{5i} = 0$, and $\epsilon^0 = \sigma |\epsilon^0|$ , with $\epsilon^0$
constant, (31) becomes
$$ M \ddot x^0 =e_0 |\epsilon^0| , \eqno(35) $$
and hence, the $dt/d\tau$ changes sign when $\tau$ becomes sufficiently large.
Note that this effect is independent of velocity.  Stueckelberg [5] found, in
his study, that it was necessary to add another, {\it ad hoc}, vector field in
order to achieve pair creation.  In fact, the structure of the Maxwell
interaction assures that the square of the proper time, as we have pointed out
above, is conserved; as seen from (12), $f^{5\mu}$ provides the possibility of
a reversal in the direction of the world line.
\par As an another example of how the Lorentz force for the pre-Maxwell case
differs from that for the Maxwell case, consider motion in a constant electric
field along the $z$-axis.  The Lorentz force for the Maxwell fields has the
form
$$ m {d^2 \over ds^2 } x^\mu = eF^{\mu\nu} {dx_\nu \over ds} \eqno(36)$$
which in the case ${\bf E} = E \hat{\bf z} $ becomes
$$ m {d \over ds }
\left( \begin{array}{c}
\dot t \\  \dot z
\end{array} \\ \right)
 = e 
\left( \begin{array}{cc}
0&E \\ E&0 \\ 
\end{array} \right)
\left( \begin{array}{c}
\dot t\\ \dot z \\
\end{array} \right)
\eqno(37)$$
Taking boundary conditions, $\dot z(0) ={dz(0) \over ds} =0$ ; $\dot t(0) =1$
; $t(0)=0$, this first order differential equation has the solution 
$$ t(s)={m \over eE}sinh {eE \over m} s \qquad z(s)={m \over eE}(cosh {eE
\over m} s -1) +z(0) \eqno(38)$$
which can be rewritten as
$$ z(t) = z(0) + {m \over eE}(\sqrt{1+(eEt/m)^2} -1) \eqno(39)$$
where we have used $dt/ds = 1/\sqrt{1-(dz/dt)^2} $ and $cosh[sinh^{-1} (x)] =
\sqrt{1+x^2} $.
\par The pre-Maxwell equivalent of such a constant (in space-time) electric
field is given by ${\bf e} = e(\tau)\hat{\bf z} $, where $\int d\tau e(\tau)
=E $.  However, from the last of eqs. (23), it is evident that a
$\tau$-dependent ${\bf e}$   must be accompanied by a space-time-dependent
$\epsilon$ field.  We may control this complication by deriving the pre-
Maxwell field from
$$ a^3 (x,\tau) = 
\begin{cases}
 -{Et \over T } , & if \mid \tau \mid  < {T \over 2};
\cr 0 , & otherwise. \cr 
\end{cases}
\eqno(40)$$
so that
$$ {\bf e} =
\begin{cases}
{E \over T } \hat{\bf z}, & if \mid \tau \mid < {T \over
2}; \cr 0, & otherwise. \cr
\end{cases}
\eqno(41)$$
$$ \int d\tau {\bf e} = E\hat{\bf z} \eqno(42)$$
and the $\epsilon$ field vanishes within the range $\mid\tau\mid < {T \over
2}$.  Now, since $f^{5\mu}$ is zero (within the range of interest), we have
from eqn. (12) that $ds/d\tau = m/M =$constant.  The equations of motion can
be reduced from
$$ M {d \over d\tau } \begin{pmatrix} dt/d\tau \\ dz/d\tau \\
\end{pmatrix} =
e_0 \begin{pmatrix}  0 & E/T \\ E/T & 0 \\ \end{pmatrix} 
\begin{pmatrix}{c} dt/d\tau \\ dz/d\tau \\ \end{pmatrix} , \eqno(43)$$
which is
$$ M {ds \over d\tau }{d \over ds } {ds \over d\tau }\begin{pmatrix} dt/ds \\ dz/ds
\\ \end{pmatrix} = e_0 \begin{pmatrix}  0 & E/T \\ E/T & 0 \\ \end{pmatrix}     {ds \over d\tau
}\begin{pmatrix} dt/ds\\ dz/ds \\ \end{pmatrix} ,$$
to
$$ m {d \over ds } \begin{pmatrix} dt/ds \\ dz/ds \\ \end{pmatrix} = e_0/T \begin{pmatrix}  0 & E \\
E & 0 \\ \end{pmatrix}     \begin{pmatrix} dt/ds \\ dz/ds \\ \end{pmatrix}. \eqno(44)$$
We therefore find that within the range $\mid\tau\mid < {T \over 2}$, the
equations of motion are the same as for the Maxwell case, with the replacement
$e \rightarrow e_0/T$, which we observe has the correct dimensionless form.
\par  We now solve the full equations of motion.   With the form of the $a^3$
field as given above, the $\epsilon$ field is given as
$$ \epsilon^3 = f^{53} = \partial^5 a^3 = -\partial_\tau [-{Et \over T}(\theta
(\tau + {T \over 2}) - \theta (\tau -{T \over 2}))] \eqno(45)$$
so that
$$ \epsilon = {Et \over T } [\delta (\tau + {T \over 2}) - \delta (\tau -{T
\over 2})] \hat{\bf z} .\eqno(46)$$
We take as initial conditions $\dot z(-\infty) =0$ and $\dot t(-\infty) =
m_0/M$.  The complete equations of motion now become
$$ M {d \over d\tau } \begin{pmatrix} dt/d\tau \\ dz/d\tau \\ \end{pmatrix} = e_0 \begin{pmatrix}  0 &
E/T \\ E/T & 0 \\ \end{pmatrix}     \begin{pmatrix} dt/d\tau \\ dz/d\tau \\ \end{pmatrix} [\theta (\tau +
{T \over 2}) - \theta (\tau -{T \over 2})] $$ 
$$ \qquad + {e_0 Et(\tau) \over T } \begin{pmatrix} 0 \\ \delta (\tau + {T \over 2})
- \delta (\tau -{T \over 2}) \\ \end{pmatrix}. \eqno(47)$$
The event will evolve freely for $\tau < -T/2$, and then encounter $\epsilon$
which gives $\dot z$ a jump to $e_0 Et(-T/2)/MT = -m_0 e_0 E/2M^2 $ and leaves
$\dot t$ unchanged.  Since $\dot x^2$ has now changed, the event transfers
mass to the $\epsilon$ field, and its new mass is
$$ m(-{T \over 2} ) = \sqrt{-M^2 (\dot x^2)} = m_0\sqrt{1-({e_0 E \over
2M})^2}, \eqno(48)$$
for $e_0 E < 2M$.  Note that if $e_0 E > 2M$, then the motion of the event
becomes {\it spacelike} (tachyonic), and the squared mass becomes negative. 
At $\tau =-{T \over 2}^+$, we find the event at $t=-m_0 T/2M$, $\dot t = m_0
/M$, $\dot z = -m_0 e_0 E/2M^2 $.  From $\tau =-{T \over 2}^+$ to $\tau = {T
\over 2}^-$, the event will evolve in the ${\bf e}$ field, according to
$$ \begin{pmatrix} \dot t(\tau) \\ \dot z(\tau) \\ \end{pmatrix} =\begin{pmatrix} cosh{e_0 E \over MT}
(\tau + {T \over 2}) & sinh{e_0 E \over MT} (\tau + {T \over 2}) \\ sinh{e_0
E \over MT} (\tau + {T \over 2}) & cosh{e_0 E \over MT} (\tau + {T \over 2})
\\ \end{pmatrix} \begin{pmatrix} \dot t(-{T \over 2}^+) \\ \dot z(-{T \over 2}^+) \\ \end{pmatrix}.
\eqno(49)$$
\vfill \eject
Then at $\tau = {T \over 2}^-$, the event will reach 
$$ t(T/2) = m_0 T [{sinh(e_0 E/M) \over e_0 E } - {cosh(e_0 E/M) \over 2M}] $$
$$ z(T/2) = z(-\infty) + m_0 T [{cosh(e_0 E/M) \over e_0 E } - {sinh(e_0 E/M)
\over 2M}] \eqno(50)$$
At $\tau = T/2$, the event will again encounter the $\epsilon$ field, where
$\dot t$ will be unchanged, while $\dot z$ will jump by 
$$\dot z({T \over 2}^+) - \dot z({T \over 2}^-)= -e_0 Et(T/2)/MT = -m_0 e_0
E/M [{sinh(e_0 E/M) \over e_0 E } - {cosh(e_0 E/M) \over 2M}], \eqno(51)$$
{\it which returns $\dot z$ to zero}.  Then for $\tau > T/2$, the event will
evolve freely, with $\dot z =0$, and 
$$\dot t = {m_0 \over M} [cosh({ e_0 E \over M}) -({e_0 E \over M})sinh({e_0 E
\over 2M})] \eqno(52)$$
which for large $e_0 E /2M$ becomes,
$$\dot t = {m_0 \over 2M} \; [1-{e_0 E \over 2M}] \; e^{({ e_0 E \over M})}
\eqno(53)$$
and we see that the condition for pair creation is $e_0 E > 2M$, which is the
condition that the evolution be spacelike when $\mid\tau\mid < {T \over 2}$.
\par   The example we have given above illustrates the striking differences
that may exist between the effects of the pre-Maxwell Lorentz forces and what
would be expected from the Lorentz forces associated with the corresponding
(through integration over $\tau$) Maxwell field.  If, on the other hand, one
considers an example with adiabatic dependence of the pre-Maxwell field, the
term $f^{35}$ can be made very small.  In this case, the pre-Maxwell Lorentz
forces result in a motion close to what would be predicted from the Maxwell
Lorentz forces (as discussed in connection with Eq. (44)).  This type of
example can be constructed by taking for $a^3$ a Gaussian function of ${\tau^2
\over T^2}$, with $T$ taken sufficiently large. 
\par To study the plane wave solutions for the source free case, we take the
Fourier transform of equations (23), resulting in
$$ {\bf k} \cdot {\bf e} =\sigma \kappa \epsilon^0 \qquad {\bf k} \cdot {\bf
h} =0 $$
$$ {\bf k} \times {\bf e} -k^0 {\bf h} =0 \qquad {\bf k} \times {\bf h} +k^0
{\bf e} =\sigma \kappa \boldsymbol{\epsilon} \eqno(54) $$
and
$$ {\bf k} \times \boldsymbol{\epsilon} -\kappa {\bf h} =0 \qquad {\bf k} \cdot \boldsymbol{\epsilon} -k^0 \epsilon^0 =0 $$
$$ -\kappa {\bf e} + k^0 \boldsymbol{\epsilon} = {\bf k} \epsilon^0 , \eqno(55) $$
where we have used the fact that $\sigma^2 =1$ and defined the transform by
$$ f(x,\tau) = \int d^4 k d\kappa e^{i(k \cdot x + \sigma \kappa \tau)}
f(k,\kappa) . \eqno(56) $$
We remark that integration of (56) over $\tau$ (or, through the Riemann-
Lebesgue lemma, in the limit $\tau \rightarrow \infty$; the static limit
referred to above corresponds to taking $\kappa =0$) selects the $\kappa =0$
component; in this case, eqs.(54) reduce to the usual Maxwell form (for which
${\bf e}$, as well as ${\bf h}$, is perpendicular to ${\bf k}$; they are
orthogonal to each other and ${\bf e}^2 = {\bf h}^2 $).  Eqs. (54) and (55)
decouple and it follows that $\boldsymbol{\epsilon}$ becomes {\it parallel} to ${\bf
k}$, $k^0 = |{\bf k}|$ and $\epsilon^0 =|\boldsymbol{\epsilon}|$.  In fact, in this
case, the two null vectors are parallel, i.e., $\epsilon^\mu = (\epsilon^0
/k^0 )k^\mu $.  With this ``natural'' limit in mind, we decompose
$$ {\bf e} = {\bf e}_{\perp} + {\bf e}_{\parallel} \qquad \boldsymbol{\epsilon} = \boldsymbol{\epsilon}_{\perp} + \boldsymbol{\epsilon}_{\parallel} \eqno(57) $$
and, in terms of the field components ${\bf e}_{\perp}$ and $\boldsymbol{\epsilon}_{\parallel}$, which we take to be independent, one finds from eqs.
(54) and (55) that
$$ {\bf e} = {\bf e}_{\perp} + \sigma ({\kappa\over k^0}) \boldsymbol{\epsilon}_{\parallel} \qquad {\bf h} = {1\over k^0 }{\bf k} \times {\bf
e}_{\perp} $$
$$ \boldsymbol{\epsilon} = \boldsymbol{\epsilon}_{\parallel} +({\kappa\over k^0}) {\bf
e}_{\perp} \qquad \epsilon^0 = {1\over k^0 }{\bf k} \cdot \boldsymbol{\epsilon}_{\parallel} $$
$$ {\bf k}^2 -(k^0 )^2 + \sigma \kappa^2 =0 . \eqno(58) $$
As in the Maxwell case, the premagnetic field ${\bf h}$ is normal to the
propagation vector ${\bf k}$; the two pre-electric fields ${\bf e}$, $\boldsymbol{\epsilon}$, are in a plane normal to ${\bf h}$, but need not be normal to
${\bf k}$.  In the Maxwell limit, as discussed above, $\kappa \rightarrow 0$,
${\bf e} \rightarrow {\bf e}_{\perp} $, ${\bf k}^2 -(k^0 )^2 \rightarrow 0$.
\par The energy-momentum tensor tensor $T^{\alpha\beta}$, which is conserved
in the absense of sources, is given [3] by,
$$ T^{\alpha\beta} = \lambda (-{1 \over 4}g^{\alpha\beta}
f^{\gamma\delta}f_{\gamma\delta} + f^{\alpha\gamma} f^{\beta}_{\gamma} ).
\eqno(59) $$
In terms of the non-covariant fields, we evaluate the following components of
$T^{\alpha\beta}$ for the plane wave solutions:
$$ T^{00} = {\lambda \over 2} [{\bf e}^2 + {\bf h}^2 +\sigma (\boldsymbol{\epsilon}^2
+ (\epsilon^0 )^2)] $$
$$ T^{0i} = \lambda [{\bf e} \times {\bf h} + \sigma \epsilon^0
\boldsymbol{\epsilon}^i ] $$
$$ T^{55} = {\lambda \over 2} [\sigma ({\bf e}^2 - {\bf h}^2 ) + \boldsymbol{\epsilon}^2 - (\epsilon^0 )^2 ] $$
$$ T^{5i} = \lambda [ \epsilon^0 {\bf e} + \boldsymbol{\epsilon} \times {\bf h} ]^i
$$
$$ T^{50} = \lambda ({\bf e} \cdot \boldsymbol{\epsilon}) . \eqno(60) $$
We recall that $T^{00}$ and $P^i = T^{0i}$ are the usual energy density and
Poynting vector (three-momentum) of the fields.  Moreover, the O(3,1)
invariant $T^{55}$ represents the mass density of the fields, while the
Poynting four-vector $S^{\mu} =T^{5\mu}$ is associated with the motion of mass
density into time and space directions.  When we evaluate these components for
the plane wave solutions, we obtain the following expressions:
$$ T^{00} = \lambda [{\bf e}^2_{\perp} + \sigma \boldsymbol{\epsilon}^2_{\parallel} ]
$$
$$ {\bf P} = T^{00} \; {\bf k} /k^0 $$
$$ T^{55} =  T^{00} \; (\kappa /k^0 )^2 $$
$$ {\bf S} = T^{00} \; ({\bf k}/k^0 )(\kappa /k^0 ) $$
$$ S^0 = T^{00} \; \kappa /k^0  \eqno(61) $$
These expressions demonstrate that the two Poynting vectors, ${\bf P }$ and
${\bf S}$ are parallel to the propagation vector ${\bf k}$.  Moreover,
$$ (T^{00} , {\bf P} ) =  ( T^{00} /k^0 ) (k^0 , {\bf k} ) $$
$$ (S^0 , {\bf S} ) = (T^{00} , {\bf P} ) (\kappa /k^0 ) \eqno(62) $$
Therefore, both $T^{0\mu}$ and $T^{5\mu}$ are in the direction of the    
four-momentum $k^\mu $.  Note that the O(3,1) invariant $\epsilon_\mu
\epsilon^\mu $ has the form $ \epsilon_\mu \epsilon^\mu = T^{55} / \lambda $. 
The factor $\kappa / k^0 $, which appears in (58), (61), and (62), is
essentially the phase velocity in the time direction of the plane wave, that
is $\Delta t / \Delta \tau $.  The relationships (61) among the components of
$T^{\alpha\beta}$ are also of a kinematical nature, and may be obtained from
the conservation of energy-momentum, the proportionality of $T^{\alpha\mu}$ to
$k^\mu$, and the five dimensional mass-shell condition $g_{\alpha\beta}
T^{5\alpha} T^{5\beta} =0$. These relationships are shown in Figure 1.
\begin{center}
\includegraphics[width=6.0in]{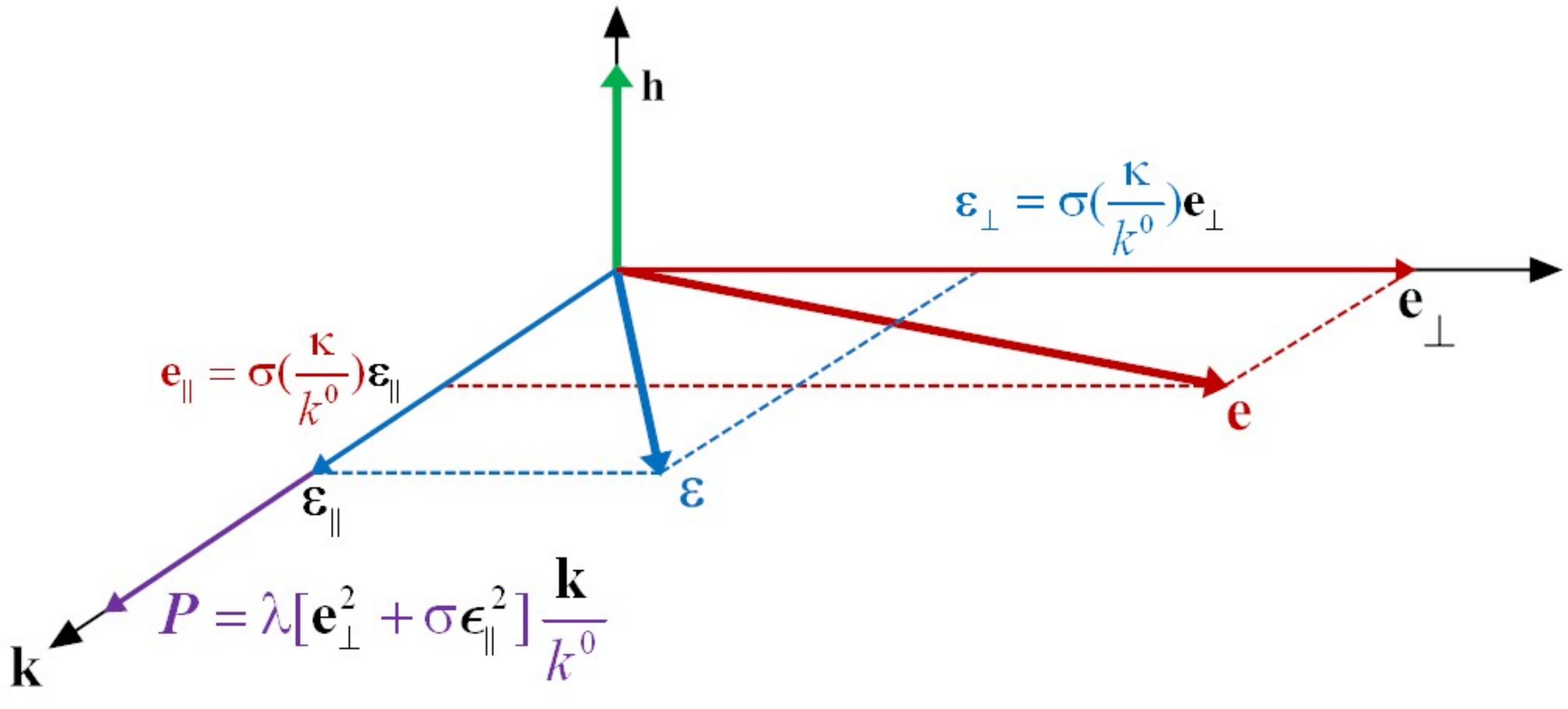}\\
\vspace{12pt}
\begin{tabular}{l}
{\normalsize \ \textbf{Figure 1}: The fields $\mathbf{e}$, $\mathbf{h}$, and
$\boldsymbol{\epsilon}$ with the Poynting vector $\mathbold{P}$.} \\ 
\end{tabular}%
\end{center}
\vspace{12pt}
\par  In [3], it is shown that in the presence of sources, the energy-momentum
tensor satisfies
$$ -\partial_\beta T^{\alpha\beta} (x,\tau) = e_0 f^{\alpha\beta} j_\beta
(x,\tau ) \eqno(63) $$
If we take for $j_\beta $, the single-particle current defined in (15),
$$ j_\beta (x,\tau ) = \dot r_\beta (\tau) \delta ^{4} (x-r(\tau )) \eqno(64)
$$
then we may integrate (63), 
$$ \int d^{4} x [\partial_\beta T^{\alpha\beta} (x,\tau) + e_0 f^{\alpha\beta}
j_\beta (x,\tau ) ]=0 $$
and, with appropriate boundary conditions for $T^{\alpha\beta}$, we obtain,
$$ [ {d \over d\tau} \int d^{4} x T^{5\alpha} (x,\tau ) ] + e_0
f^{\alpha\beta} (r,\tau ) \dot r_\beta (\tau ) =0 . \eqno(65) $$
We recognize the second term in (65) as the covariant form of the Lorentz
force, and we may use (13) to rewrite (65) as
$$ {d \over d\tau} \{ \int d^{4} x T^{5\alpha} (x,\tau ) + M [p^\alpha (\tau )
-e_0 a^\alpha (r(\tau ), \tau ) ] \} =0, \eqno(66) $$
which expresses the conservation of the total energy-momentum-mass of the
particle and fields.
\newpage
\par{\bf References}
\smallskip
\begin{enumerate}
\item E.C.G. Stueckelberg, Helv.\  Phys.\  Acta\  {\bf 14}, 322 (1941);
{\bf 14}, 588 (1941).
\item L.P. Horwitz and C. Piron, Helv.\  Phys.\  Acta\  {\bf 48}, 316
(1973).
\item D. Saad, L.P. Horwitz and R.I. Arshansky, Found.\  of\  Phys.\ {\bf
19}, 1125 (1989). See also, L. Hostler, Jour.\ Math.\ Phys.\ {\bf 21},2461
(1979); {\bf 22}, 2307 (1980), and R. Kubo, Nuovo Cimento {\bf 85A}, 4(1984).
\item M.C. Land and L.P. Horwitz, Found.\  of\  Phys.\ , to appear (1990).
\item E.C.G. Stueckelberg, Helv.\  Phys.\  Acta\ {\bf 15}, 23 (1942).
\item R.I. Arshansky, L.P. Horwitz and Y. Lavie, Found.\  of\  Phys.\ {\bf
13}, 1167 (1983).
\end{enumerate}

\end{document}